\documentstyle [12pt,epsf] {article}

\catcode`@=12
\newcommand{\beq}{\begin{equation}}
\newcommand{\eeq}{\end{equation}}
\newcommand{\bea}{\begin{eqnarray}}
\newcommand{\eea}{\end{eqnarray}}
\newcommand{\bers}{\begin{eqnarray*}}
\newcommand{\eers}{\end{eqnarray*}}
\newcommand{\nn}{\nonumber}
\newcommand{\bt}{\begin{itemize}}
\newcommand{\et}{\end{itemize}}

\begin{document}
\vspace{0.5in}
\oddsidemargin -.375in
\newcount\sectionnumber
\sectionnumber=0
\def\bra#1{\left\langle #1\right|}
\def\ket#1{\left| #1\right\rangle}
\def\be{\begin{equation}}
\def\ee{\end{equation}}
\def\lsim{\mathrel{\vcenter{\hbox{$<$}\nointerlineskip\hbox{$\sim$}}}}
\thispagestyle{empty}
%\begin{flushright}
%UH-PHYS-TH-03-15\\
%\end{flushright}
\vskip0.5truecm

\begin{center}

{\large \bf
\centerline{New physics effects on the CP asymmetries in }} 
{\large \bf \centerline { \boldmath{$B \rightarrow \phi K_S$} and 
\boldmath{$B \rightarrow \eta' K_S$} decays}}
\vspace*{1.0cm}
%\vskip1cm
{ Anjan K. Giri$^1$ and  Rukmani Mohanta$^2$ } \vskip0.3cm
{\it  $^1$ Physics Department, Technion-Israel 
Institute of Technology, 32000 Haifa, Israel}\\
{\it $^2$ School of Physics, University of Hyderabad,
Hyderabad - 500046, India
} \\
\vskip0.5cm
\bigskip
(\today)
\vskip0.5cm

\begin{abstract}
Within the standard model (SM), the time dependent CP asymmetries in
$B \to \psi K_S$, $ B \to \eta' K_s$ and $B \to \phi K_s$
are expected to give the same result i.e. $\sin 2 \beta $. However, 
recent measurements of the mixing induced CP asymmetries in $B
\rightarrow \phi K_S$ and $B \to \eta' K_s$ modes give results 
whose central values differ
from the SM expectations. We explore the effect of new
physics in the two Higgs doublet model (THDM), which allows
tree level flavor changing neutral currents (so called Model
III), and the model with extra vector-like down quark (VLDQ).
We find that the observed mixing induced CP
asymmetry for $B \to \phi K_S$ can not be accommodated by the THDM, 
but can be
explained in the VLDQ model and both models can explain the observed 
asymmetry for $B \to \eta' K_S$ mode.
\end{abstract}
\end{center}
\newpage
\baselineskip=14pt

\section{Introduction}

A new era in B-physics has just been started with the advent of
$B$-factories. With the accumulation of huge data in the B-
system, the standard model (SM) will be subjected to a very
stringent test. At the same it is also considered that the
experiments at B factories are also potential sources for probing
new physics. The BaBar \cite{babar} and Belle \cite{belle}
measurements of the time-dependent asymmetries in the gold plated
mode $B\to \psi K_S$ have provided the first evidence of CP
violation in the $B$-system. The observed world average of $\sin
2\beta$ \cite{nir}, 
\bea \sin 2\beta_{\psi K_S}=0.734\pm 0.054,
\label{beta} \eea 
agrees well with the SM prediction. This
indicates that CP symmetry is significantly violated in nature and
the Kobayashi-Maskawa (KM) mechanism \cite{km} seems to be the
dominant source of CP violation, in which the phase $\delta_{KM}$
is the only source of CP violation. However, this speculation does
not exclude interesting CP violating new physics (NP) effects in
other $B$ decays. Since the decay $B\to \psi K_S$ $(b\to c\bar c
s)$ is a tree level process in the SM, the NP contributions to its
amplitude are naturally suppressed. Moreover, at loop level NP may
give large contributions to the $B^0$-$\bar B^0$  mixing as well
as to the loop-induced decay amplitudes.  The former effects are
universal to all $B^0$ decay modes and are constrained to be
less than 20\% compared to that of the SM contribution \cite{nir}.
On the other hand, the effects of new physics in the decay
amplitudes are non-universal, may vary from process to process and
can show up in the comparison of the CP asymmetries in different
decay modes \cite{gross}.

One of the most promising processes for NP searches widely
considered in literature [6-14] is the  decay $B \to \phi K_S$.
Various NP scenarios have been presented to explain the data.
Unfortunately, we do not know at present which is the correct one. 
Hopefully, careful study in future will rule out some of the
scenarios, at least as far as the understanding of B physics and CP
violation is concerned.

Unlike $B \to \psi K_S$, the process $B \to \phi K_S$ has no tree
level amplitude, which makes inroads for NP to play an important
role in this mode. In the SM the decay $ b \to s \bar s s $, which
contributes to $B\to \phi K_S$, is induced at one loop level.
Thus, it is natural to expect that new physics contribution to
this decay mode may be quite significant. According to the KM
mechanism of CP violation, both CP asymmetries in $B \to \phi K_S$
and $B \to \psi K_S$ processes should measure the same quantity,
namely $\sin 2\beta$, with negligible hadronic uncertainties (upto
${\cal O}(\lambda^2)$, $\lambda\approx 0.2$) \cite{gross,gr1}.
However, contrary to the SM expectations, the recent measurements
of CP asymmetries in $B \to \phi K_S$ by BaBar \cite{babar1} and
Belle \cite{belle} collaborations have registered  significant
deviation from the predictions, as

\begin{eqnarray}
\sin (2 \beta)_{\phi K_S}&=&-0.19 ^{+0.52}_{-0.50} \pm 0.09
~~~~~~~{\rm BaBar}\nn\\
\sin (2 \beta)_{\phi K_S}&=&-0.73 \pm 0.64 \pm 0.18 ~~~~{\rm
Belle}
\end{eqnarray}
with an average
\beq \sin ((2 \beta)_{\phi K_S})_{ave}=-0.39 \pm
0.41
\eeq
The corresponding branching ratio is given (in units of
$10^{-6}$) as
\begin{eqnarray}
BR(B \to \phi K^0)&=&8.7^{+1.7}_{-1.5} \pm 0.9
~~~~~~{\rm BaBar}\nn\\
BR(B \to \phi K^0)&=& 10.0^{+1.9+0.9}_{-1.7-1.3}
~~~~~~~{\rm Belle}\nn\\
BR(B \to \phi K^0)&=&5.4^{+3.7}_{-2.7} \pm 0.7 ~~~~~~{\rm CLEO}~
\cite{cleo}
\end{eqnarray}
with an average
\beq BR(B \to \phi K^0)_{ave}=8.67 \pm
1.28\;.\label{eq:br1}
\eeq

One can see that there are large statistical errors associated
with these measurements. Nevertheless, the data establish a $2.7
\sigma$ deviation from the SM prediction $\sin(2 \beta)_{\phi
K_S}=\sin (2 \beta)_{\psi K_S}$. Therefore, if the measurement of
$\sin 2\beta$ in $ B\to \psi K_S$ is considered as the first
evidence of large CP violation in B system then the difference
between $\sin(2 \beta)_{\phi K_S}$ and $\sin (2 \beta)_{\psi K_S}$
is likely to be regarded as a potential hint for the presence of
new physics. There are several attempts in the literature [6-14]
with detail discussion on the possible implications of this
result.

 The second channel we are interested in is $B^0 \to \eta' K_S$. This
is another two-body decay mode which is similar to the two mentioned above.
Since many alternative schemes have been presented in the literature 
recently to explain the $\sin (2\beta)_{\phi K_s}$ deviation, 
it is therefore very 
important to verify that each of the NP scenarios should successfully
explain them all. At present it is difficult to say which is the correct
description. In order to narrow down the same it is highly desirable
that one should carefully study them. This will not only help us to narrow 
down the sources of NP but also provide important clues for hadronic 
B-physics in general.

$B^0 \to \eta' K_S$
also receives dominant contribution from the 
$ b \to s \bar s s $ gluonic penguin, and therefore it is expected that the 
time dependent mixing induced CP asymmetry for this 
mode will also give the value
$\sin 2 \beta $ \cite{nir}. However, this decay mode also has a 
tiny CKM as well as
color suppressed $b \to u \bar u s$ tree contributions along with
$b \to s \bar q q~(q=u,d)$ penguins, which induce deviation from the
leading result. It has been shown in Ref. \cite{lon} that
this deviation will be below two percent level. Belle \cite{abe} 
and BaBar \cite{bab} collaborations have 
recently measured the CP asymmetry for this mode which is given as
\begin{eqnarray}
\sin (2 \beta)_{\eta' K_S}&=&0.71 \pm 0.37 ^{+0.05}_{-0.06} 
~~~~~~~{\rm Belle}\nn\\
\sin (2 \beta)_{\eta' K_S}&=&0.02 \pm 0.34 \pm 0.03 ~~~~{\rm
BABAR}
\end{eqnarray}
with an average
\beq \sin ((2 \beta)_{\eta' K_S})_{ave}=0.33 \pm
0.25\;,
\eeq
whose central value also deviates significantly from SM
expectations.

In this paper we would like to investigate the new physics effects
on the CP asymmetry parameters of the decay $B^0 \to \phi K_S$ and $B^0
\to \eta' K_S$ modes, 
arising from some simple extensions of the SM. The models
considered here are the two Higgs doublet models (THDM) which
allows tree level flavor changing neutral currents, the so called
model III and the model with extra vector-like down quarks (VLDQ).
We show that the observed data for $B^0 \to \phi K_S$ 
can be easily accommodated  in the
VLDQ model whereas it can not be explained in the THDM, and
both the models can explain the data for $B^0 \to \eta' K_S$ mode.
It has already been discussed in Ref.\cite{ref5}, that whether
these two models
can explain the observed CP asymmetry in $ B \to \phi K_S$ mode, i.e.
$\sin (2 \beta )_{\phi K_S}$. However in this paper we have explicitly done 
the calculation for both the decay modes and confirm the result of Ref.
\cite{ref5} for the decay mode $ B \to \phi K_S$.

The paper is organized as follows. In section II, we present the basic 
formulae for CP violating parameters, in the presence of new physics.
In section III, we discuss CP violation effects in $B^0 \to \phi K_S$ mode
arising from the THDM and VLDQ model. The $B^0 \to \eta' K_S$
process is discussed in section IV. Section V contains our conclusion.

\section{CP violation parameters}
Here, we will present the basic formulae of CP asymmetry
parameters, in the presence of new physics. Due to the contributions 
from new physics, these
parameters deviate substantially from their standard model values.
Let us consider the $B^0$ and ${\bar B}^0$ decay into a
CP eigenstate $f_{CP}$ (we consider $f_{CP}=\phi K_S $ or $\eta'K_S$).
Here, we are presenting the formulae for $B^0 \to \phi K_S$ mode, but the same
results will also hold for $B^0 \to \eta' K_S$ mode.
The time dependent CP asymmetry for $B \to \phi K_S$
can be described by \cite{cp}
\beq
{\cal A}_{\phi K_S}(t)=C_{\phi K_S}\cos (\Delta M_{B_d} t)+ S_{\phi K_S}
\sin (\Delta M_{B_d} t)
\eeq
where we identify
\beq
C_{\phi K_S}=\frac{1-|\lambda|^2}{1+|\lambda|^2},~~~~
    S_{\phi K_S}=-\frac{2 {\rm Im}(\lambda)}{1+|\lambda|^2}\;,
\eeq
as the direct and the mixing-induced CP asymmetries.
The parameter $\lambda$ corresponds to
\beq
\lambda=\frac{q}{p}\frac{A(\bar B \to \phi K_S)}{A(B \to \phi K_S)}
\eeq
where, $q$ and $p$ are the mixing parameters  and
represented by the CKM
elements in the standard model as
\beq
\frac{q}{p}=\frac{V_{tb}^* V_{td}}{V_{tb} V_{td}^*}
\eeq
Using CKM unitarity, the amplitude for $ \bar B \to \phi K_S$ 
is given as \cite{gr1,gr2}
\beq
A(\bar B \to \phi K_S)=\lambda_c A^{cs}+\lambda_u A^{us}
\eeq
where $\lambda_q= V_{qb}V_{qs}^*$. The first term which is the dominant one,
is real. Thus if one neglects the subdominant amplitude i.e. the doubly 
Cabibbo supressed second term which in general expected to be very small,
the mixing induced CP asymmetry 
is given as,
$S_{\phi K_S}=\sin ~2 \beta $,
same as the one for $B \to \psi K_S$ in the SM. It has
beeen shown in Ref. \cite{gr1}, that the correction 
due to the second term is upto ${\cal O}(\lambda^2)$ i.e.,
\beq
|S_{\phi K_S}-\sin 2 \beta | \leq {\cal O}(\lambda^2)\;.
\eeq
Adding a mild dynamical assumption to the SU(3) analysis, recently it
has been shown in Ref. \cite{gr2} that the upper bound
of standard model 
pollution to the dominant amplitude of $B \to \phi K_S $ mode 
is of the order of 0.25
and for $B \to \eta' K_S$ as 0.3.

New Physics could in principle contribute to both mixing and decay
amplitudes. The new physics contribution to mixing is universal
while it is non universal and process dependent in the decay
amplitudes. As the NP contributions to mixing phenomena is
universal, it will still set $ S_{\psi K_S}=S_{\phi K_S}$.
Therefore, to explain the observed $2.7 \sigma $ deviation in $
(S_{\psi K_S}-S_{\phi K_S})$, here we explore the NP effects only
in the decay amplitudes. Thus including the NP contributions, we
can write the decay amplitude for $ B \to \phi K$ process as

\beq
A(B^0 \to \phi K)=A_{SM}+A_{NP}= A_{SM}\left [ 1+ r_{NP}~ e^{i
\phi_{NP}} \right ]
\eeq
where $r_{NP}=|A_{NP}/A_{SM}|$, ($A_{SM}$
and $A_{NP}$ correspond to the SM and NP contributions to the $B
\to \phi K_S$ decay amplitude) and $\phi_{NP}={\rm
Arg}(A_{NP}/A_{SM} )$, which contains both strong and weak phase
components.

The branching ratio for $B \to \phi K$ decay process can be given
as
\beq
BR(B \to \phi K)=BR^{SM}\left ( 1+r_{NP}^2 +2 r_{NP} \cos
\phi_{NP} \right )
\eeq
where $BR^{SM}$ represents the
corresponding standard model value.

Now if we write $\phi_{NP}=\delta_{NP}+\theta_{NP}$, where
$\delta_{NP}$ and $\theta_{NP}$ are the relative strong and weak
phases between the new physics contributions to the decay
amplitude and the SM part, one can then obtain the expressions for
the CP asymmetries as
\beq S_{\phi K} =\frac{\sin 2 \beta+ 2
r_{NP} \cos \delta_{NP} \sin(2\beta +\theta_{NP})+r_{NP}^2 \sin (2
\beta+2 \theta_{NP})}{ 1+r_{NP}^2 +2 r_{NP} \cos \delta_{NP}\cos
\theta_{NP}}\label{eq:eq1}
 \eeq
 and
 \beq
 C_{\phi K} =\frac{-2
r_{NP} \sin \delta_{NP} \sin \theta_{NP}}{ 1+r_{NP}^2 +2 r_{NP}
\cos \delta_{NP}\cos \theta_{NP}}\label{eq:eq2}
 \eeq

In Eqs. (\ref{eq:eq1}) and (\ref{eq:eq2}) there are three
unknowns, namely, $r_{NP}$, $\theta_{NP}$ and $\delta_{NP}$. So if
somehow we could constrain the value of $r_{NP}$ considering
different new physics models, we could vary the $\theta_{NP}$ and
$\delta_{NP}$ parameters to obtain the required value of $S_{\phi
K}$. 

\section{CP Violation in $B \to \phi K_S$ process} 

To study the CP violation effects in $B^0 \to \phi K_S$
process, first we present the SM amplitude and
then we consider the THDM and thereafter the model with extra
vector-like down quark, in the following subsections.

\subsection{SM contributions}
In the SM, the decay process $B \to \phi K_S$ proceeds through the
quark level transition $b \to s \bar s s$, which is induced by the
QCD, electroweak and magnetic penguins. QCD penguins with the top
quark in the loop contribute predominantly to such process.
However, since we are looking for NP here we would like to retain
all the contributions. The effective Hamiltonian describing the
decay $b\to s\bar ss$ \cite{flei1} is given as
\beq 
H_{eff}=
-\frac{G_F}{\sqrt{2}}V_{tb}V_{ts}^* \left( \sum_{j=3}^{10}C_j O_j
+C_g O_g \right),
\eeq
where $O_3, \cdots, O_{6}$ and $O_7,
\cdots, O_{10}$ are the standard QCD and EW penguin operators,
respectively, and $O_g$ is the gluonic magnetic operator. Within
the  SM and at scale $M_W$, the Wilson coefficients $C_1(M_W),
\cdots , C_{10}(M_W)$ at next to leading logarithmic order (NLO)
and $C_g(M_W) $ at leading logarithmic order (LO) have been given
in Ref \cite{flei2}. The corresponding QCD corrected values at the energy
scale $\mu=m_b$, can be obtained using the renormalization group
equation, as described in Ref. \cite{lu}.

To calculate the $B$ meson decay rate, we use the factorization
approximation to evaluate the hadronic matrix element $\langle O_i
\rangle\equiv\langle \bar K^0\phi|O_i|\bar B^0 \rangle$. Since the hadronic
matrix elements do not appear in the expressions for CP asymmetry
parameters, they will not introduce any uncertainties in the
results. In this approximation the matrix elements are given as
$\langle O_3 \rangle=  \langle O_4\rangle =4 X/3$, $\langle
O_5\rangle=X$, $\langle O_6\rangle=X/3$, $\langle O_7 \rangle =
-X/2$, $\langle O_8\rangle=-X/6$ and $\langle O_9 \rangle=\langle
O_{10}\rangle=-2X/3$, where the factorizable hadronic matrix
element $X$ is given as $X=\langle \bar K^0 (p_K)| \bar s
\gamma_\mu(1-\gamma_5)b |\bar B^0(p_B) \rangle
\langle \phi(q, \epsilon_\phi )|\bar s
\gamma^\mu(1-\gamma_5)s|0 \rangle = 2 F^{BK}_1(M_\phi^2)f_\phi M_\phi
\epsilon_\phi \cdot p_K$.  For evaluating the matrix element of
the most relevant operator, i.e., $O_g$, we use the procedure of
\cite{rid}, where it has been shown that the operator $O_g$ is
related to the matrix element of the QCD and electroweak penguin
operators as
\bea
 \langle O_g\rangle &=&-\frac{\alpha_s}{4\pi}\frac{m_b}{\sqrt{
 \langle q^2 \rangle }}\left [
 \langle O_4\rangle + \langle O_6\rangle
 -\frac{1}{N_C}(\langle O_3\rangle+\langle O_5\rangle)\right ]
\eea
where $q^\mu$ is the momentum transferred by the gluon to the
$(\bar s,s)$ pair. The parameter $\langle q^2\rangle$ introduces
certain uncertainty into the calculation. In the literature its
value is taken in the range $1/4 \lsim \langle q^2 \rangle/m_b^2
\lsim 1/2$ \cite{hou}, and we will use $\langle q^2 \rangle/m_b^2
= 1/2$ \cite{lu}, in our numerical calculations.

Thus, in the factorization approach the amplitude $A\equiv\langle
\phi K^0 |H_{eff}|B^0 \rangle$ of the decay  $B^0\to \phi K^0$ takes a form
\bea
A(\bar B^0 \to \phi \bar K^0)=-\frac{G_F}{\sqrt{2}}V_{tb}V_{ts}^* \left[
a_3+a_4+a_5-\frac{1}{2}(a_7+a_9+a_{10}) \right] X\;,
 \label{A}
 \eea
where $X$ stands for the factorizable hadronic matrix element of
which exact form is irrelevant for us since it cancels out in the
CP asymmetries. The coefficients $a_i$ are given by
\bea a_{2i-1}=
C^{eff}_{2i-1}+\frac{1}{N_c}C^{eff}_{2i}\,, \qquad\qquad a_{2i}=
C^{eff}_{2i}+\frac{1}{N_c}C^{eff}_{2i-1}\,,
\eea
where $N_C$ is
the number of colors. The values of the QCD improved effective
coefficients $C^{eff}_i$ can be found in \cite{lu,thdm2}. Now
substituting the values of $a_i$ for $N_C$=3, from \cite{thdm2},
the value of the form factor $F_1^{BK}(M_\phi^2)=$ 0.39 and using
the $\phi$ meson decay constant $f_{\phi}=$ 0.233 GeV and $\tau_{B^0}=
1.542 \times 10^{-12}$ sec \cite{pdg}, we obtain
the branching ratio in SM as
\beq
BR^{SM}(B \to \phi K^0)=10.5
\times 10^{-6}
\eeq
which lies within the present experimental
limits (\ref{eq:br1}).

\subsection{Two Higgs Doublet Model contributions}

We now proceed to calculate the new physics effect in two Higgs
doublet Model (THDM), which is one of the simplest extensions of
the SM \cite{higgs}. In such models, the tree level flavor
changing neutral currents (FCNC's) are prevented by imposing one
ad hoc discrete symmetry to constrain the THDM scalar potential
and Yukawa Lagrangian and thus one obtains
 the so called model I and
model II \cite{thdm}. In model I both the up and down type quarks get mass
from the Yukawa couplings to the same Higgs doublet $\phi_1$ and
in Model II the up- and down type quarks get their masses from
Yukawa couplings to two different scalar doublets $\phi_1$ and
$\phi_2$. Here we consider the model III \cite{thdm1} of THDM
where no discrete symmetry is imposed and both up- and down type
quarks may have diagonal and/or off diagonal flavor changing
couplings with  the two Higgs doublets $\phi_1$ and $\phi_2$.

The Yukawa Lagrangian of the quarks in model III is given in the
form \cite{thdm2}
\beq
 {\cal L}_Y^{III}=\eta_{ij}^U \bar Q_{i,L} \tilde \phi_1 U_{j, R}
+\eta_{ij}^D \bar Q_{i,L} \phi_1 D_{j, R} +\hat \xi_{ij}^U \bar
Q_{i,L} \tilde \phi_2 U_{j, R} +\hat \xi_{ij}^D \bar Q_{i,L}
\phi_2 D_{j, R}+{\mbox h.c}
\eeq
where $\phi_i~(i=1,2)$ are the
two Higgs doublets of THDM, $\tilde \phi_i=i\tau_2 \phi_i^*$,
$Q_{i,L}$ with ($i=1,2,3)$ are the left handed isodoublet quarks
$U_{j, R}~ (D_{j, R})$ are the right handed isosinglet up (down)
type quarks. $\eta_{i,j}^{U,D}$
correspond to the diagonal mass matrices of the up and down
quarks, while the neutral and charged flavor changing couplings
are
\beq
\xi_{ij}^{U,D}=\frac{ \sqrt{m_i m_j}}{v}\lambda_{ij}, ~~~
\hat \xi_{neutral}^{U,D}=\xi^{U,D}, ~~~\hat
\xi_{charged}^{U}=\xi^{U}V_{CKM}, ~~~ \hat
\xi_{neutral}^{U}=V_{CKM}\xi^D\;,
\eeq
where $V_{CKM}$ is the
Cabibbo-Kobayashi-Maskawa mixing matrix \cite{km}. The coupling
constants $\lambda_{ij}$ are the free parameters of the model to
be determined from experimental data. 

Recently Chao et al
\cite{cck} studied the $b \to s \gamma$ process and Xiao et al
\cite{thdm2} studied the charmless nonleptonic decays of $B$
mesons using the model III of THDM where they have kept only the couplings
$\lambda_{tt}=|\lambda_{tt}|e^{i \theta_t}$ and $\lambda_{bb}=
|\lambda_{bb}|e^{i \theta_b}$ as nonzero. From the studies
of \cite{thdm2,cck}, it is known that, the following parameter space for
 model III 
\bea
&& \lambda_{ij}=0 ~~{\rm for}~ij \neq tt~~{\rm or}~~bb \nn \\
&&  |\lambda_{tt}|=0.3,~~|\lambda_{bb}|=35,~~ \theta=(0^\circ -
30^\circ), ~~M_{H^+}=(200 \pm 100)~{\rm GeV}\,
 \eea
where $\theta=\theta_b -\theta_t$, are allowed by the
available data. The advantage of
keeping only these two couplings nonzero is that the neutral Higgs
boson do not contribute at the tree level or one loop level. The
new contributions therefore come only from the charged Higgs
penguin loop with heavy internal top quark.

The new physics will manifest itself by modifying the
corresponding Inami-Lim  \cite{lim} functions $C_0(x)$, $D_0(x)$,
$E_0(x)$ and $E_0'(x)$ which determine the Wilson coefficients
$C_3(M_W), \cdots , C_{10}(M_W)$ and $C_g(M_W)$ in SM. The new
strong and electroweak penguin diagrams in THDM can be obtained
from the corresponding penguin diagrams in SM by replacing the
internal $W^\pm$ lines by the charged Higgs $H^\pm$ lines.
Following the same procedure as in the SM, it is straight forward
to calculate the new $\gamma$-, $Z^0$ and gluonic penguin diagrams
induced by the exchange of charged Higgs bosons in Model III.
These new Wilson coefficients $C_i^{H^\pm}(M_W)$ $i=3, \cdots, 10$
at NLO level and $C_g$ at the LO level can now be written as

\bea
C_3^{H^\pm}(M_W)&=& -\frac{\alpha_s(M_W)}{24 \pi} E_0^{NP}
+\frac{\alpha_{em}}{6 \pi} \frac{1}{\sin^2 \theta_W} C_0^{NP}\nn\\
C_4^{H^\pm}(M_W)&=& \frac{\alpha_s(M_W)}{8 \pi} E_0^{NP}
\nn\\
C_5^{H^\pm}(M_W)&=& -\frac{\alpha_s(M_W)}{24 \pi} E_0^{NP}
\nn\\
C_6^{H^\pm}(M_W)&=& \frac{\alpha_s(M_W)}{8 \pi} E_0^{NP}
\nn\\
C_7^{H^\pm}(M_W)&=& \frac{\alpha_{em}}{6 \pi}
\left [4 C_0^{NP}+D_0^{NP}\right ]
\nn\\
C_8^{H^\pm}(M_W)&=&C_{10}^{H^\pm}(M_W)=0\nn\\
C_9^{H^\pm}(M_W)&=& \frac{\alpha_{em}}{6 \pi}
\left [4 C_0^{NP}+D_0^{NP} + \frac{1}{\sin^2 \theta_W}
4 C_0^{NP} \right ]
\nn\\
C_g^{H^\pm}(M_W)&=& -\frac{1}{2} E_0^{'{NP}}
\eea
where the
functions $C_0^{NP},$ $D_0^{NP}$, $E_0^{NP}$ and $E_0'^{NP}$ are
the new physics contributions to the Wilson coefficients
 arising from the charged Higgs exchange
penguin diagrams. These are given by
\bea
C_0^{NP}&=&-\frac{x_t}{16}\left [ \frac{y_t}{1-y_t}
+\frac{y_t}{(1-y_t)^2}\ln y_t \right ]
|\lambda_{tt}|^2\nn\\
D_0^{NP}&=&-\frac{1}{3} H(y_t)
|\lambda_{tt}|^2\nn\\
E_0^{NP}&=&-\frac{1}{2} I(y_t)
|\lambda_{tt}|^2\nn\\
E_0^{'{NP}}&=&\frac{1}{6} J(y_t)
|\lambda_{tt}|^2-K(y_t)
|\lambda_{tt}\lambda_{bb}|e^{i \theta}
\eea
with
\bea
H(y)&=&\frac{38 y-79 y^2 +47y^3}{72(1-y)^3}
+ \frac{4y -6y^2+3y^4}{12 (1-y)^4} \ln y \nn\\
I(y)&=&\frac{16 y-29 y^2 +7y^3}{36(1-y)^3}
+ \frac{2y -3y^2}{6 (1-y)^4} \ln y \nn\\
J(y)&=&\frac{2 y+5 y^2 -y^3}{4(1-y)^3}
+ \frac{3y^2}{2 (1-y)^4} \ln y \nn\\
K(y)&=&\frac{-3 y+y^2 }{4(1-y)^2} - \frac{y}{2 (1-y)^3} \ln y.
\eea
In the above use has been made of $x_t =m_t^2/M_W^2$ and
$y_t=m_t^2/M_{H^+}^2$.

Since the charged Higgs bosons appeared in Model III have been
integrated out at the scale $M_W$, the QCD running of Wilson
coefficients $C_i^{H^\pm}(M_W)$ down to the scale $\mu = {\cal
O}(m_b)$ using the renormalization group equation can be done in
the same way as in the SM. Including the new physics contributions the
values of the effective Wilson coefficients at the scale ${\cal
O}(m_b)$ are explicitly given in Ref. \cite{thdm2}.
Using the values for the Wilson coefficients from \cite{thdm2}, we
obtain the $B \to \phi K^0 $ amplitude in THDM as

\beq
A^{THDM}(B \to \phi K^0)=A^{SM}\left (1+0.28~ e^{i
(\theta_{NP}+\delta_{NP})} \right )\;.
 \eeq
 Now taking
$r_{NP}=0.28$ and varying the weak phase $\theta_{NP}=\{-\pi,\pi\}
$ and strong phase  $\delta_{NP}=\{0,2\pi\} $ according to Eq.
(16), we find that the
value of $S_{\phi K}$ can not be negative as shown in Fig-1.
Thus the observed value of $S_{\phi K}$ can not be accommodated in
the THDM.
%%%%%%%%%%%%%%%%%%%%%%%%%%%%%%%%%%%%%%%%
\begin{figure}[htb]
   \centerline{\epsfysize 4.0 truein \epsfbox{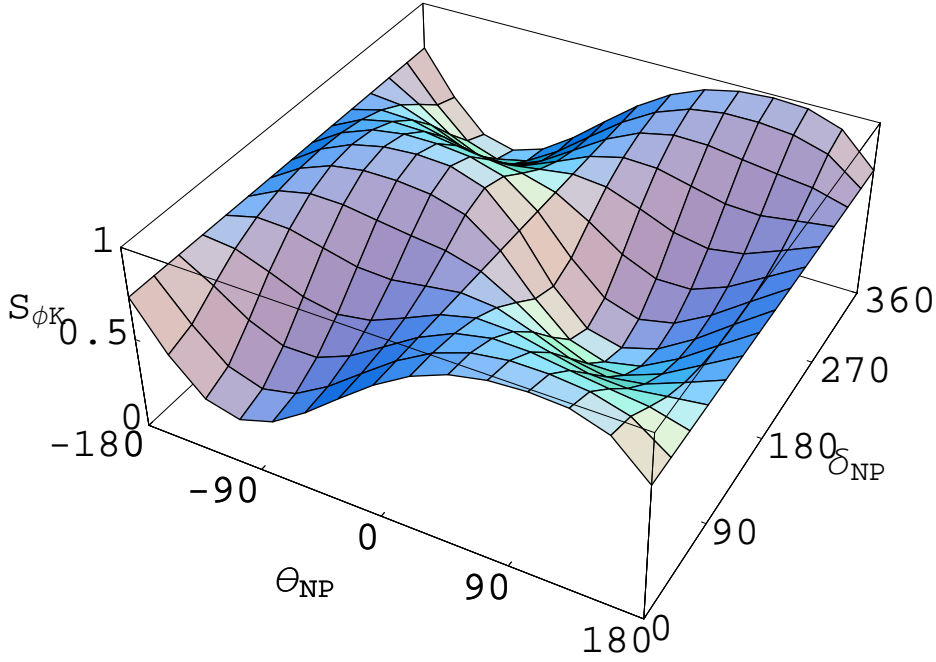}}
   \caption{
 3DPlot of $S_{\phi K}$ versus the weak phase
$\theta_{NP}$ and strong phase $\delta_{NP}$ (in degrees) for $r_{NP}=0.28$}
 \label{phiK.ps}
 \end{figure}
%%%%%%%%%%%%%%%%%%%%%%%%%%%%%%%%%%%%%%%%%%%%%%%%%%%%%%%

\subsection{Contributions from Model with extra vector like down quark}

Now we consider the model with an additional vector like down
quark \cite{ref11}. It is a simple model beyond the SM with an
enlarged matter sector with an additional vector-like down quark
$D_4$. The most interesting effects in this model concern CP
asymmetries in neutral B decays into final CP eigenstates. At a
more phenomenological level, models with isosinglet quarks provide
the simplest self consistent framework to study deviations of $3
\times 3$ unitarity of the CKM matrix as well as, allow flavor
changing neutral currents at the tree level. The presence of an
additional down quark implies a $4 \times 4$ matrix $V_{i \alpha}$
$(i=u,c,t,4,~\alpha= d,s,b,b')$, diagonalizing the down quark mass
matrix. For our purpose, the relevant information for the low
energy physics is encoded in the extended mixing matrix. The
charged currents are unchanged except the $V_{CKM}$ is now the $3
\times 4$  upper submatix of $V$. However, the distinctive feature
of this model is that FCNC enters neutral current Lagrangian of
the left handed downquarks :
 \beq
 {\cal L}_Z= \frac{g}{2 \cos
\theta_W} \left [ \bar u_{Li} \gamma^{\mu} u_{Li} - \bar d_{L
\alpha}U_{\alpha \beta} \gamma^\mu d_{L \beta}-2 \sin^2 \theta_W
J_{em}^\mu \right ]Z_{\mu}
\eeq
with
\beq
U_{\alpha \beta}
=\sum_{i=u,c,t} V_{\alpha i}^\dagger V_{i \beta} =\delta_{\alpha
\beta} - V_{4 \alpha}^* V_{4 \beta}
\eeq
where $U$ is the neutral
current mixing matrix for the down sector which is given above. As
$V$ is not unitary, $U \neq {\bf{1}}$. In particular its
non-diagonal elements do not vanish :
\beq
U_{\alpha \beta}= -V_{4
\alpha}^* V_{4 \beta} \neq 0~~~{\rm for}~ \alpha \neq \beta
\eeq

Since the various $U_{\alpha \beta}$ are nonvanishing they would
signal new physics and the presence of FCNC at tree level, this
can substantially modify the predictions for CP asymmetries. The
new element $U_{s b}$ which is relevant to our study is given as
\beq
U_{sb}= V_{us}^* V_{ub}+V_{cs}^*V_{cb}+V_{ts}^*V_{tb}
\eeq

The decay mode $B^0 \to \phi K_S$ receives the new contributions both from 
color allowed and color suppressed $Z$-mediated FCNC transitions.
The new additional operators are given as
\bea
&&O_1^{Z-FCNC}=[\bar s_\alpha \gamma^\mu (1-\gamma_5) b_\alpha][\bar s_\beta
\gamma_\mu(C_V^s -C_A^s \gamma_5) s_\beta]\nn\\
&&O_2^{Z-FCNC}=[\bar s_\beta \gamma^\mu (1-\gamma_5) b_\alpha][\bar s_\alpha
\gamma_\mu(C_V^s -C_A^s \gamma_5) s_\beta]
\eea
where $C_V^s$ and $C_A^s$ are the vector and axial vector $Z s \bar s$
couplings. Using Fierz transformation and the identity
$(C_V^s-C_A^s \gamma_5)=[(C_V^s+C_A^s)(1- \gamma_5)+(C_V^s-C_A^s)(1+ \gamma_5)]
/2$, the matrix elements of the operators are given as
\bea
\langle \phi \bar K^0 |O_1^{Z-FCNC}| \bar B^0 \rangle
&=& \left [ \frac{4}{3}\frac{(C_V^s+C_A^s)}{2}+
\frac{(C_V^s-C_A^s)}{2} \right ] X\nn\\
\langle \phi \bar K^0 |O_2^{Z-FCNC}| \bar B^0 \rangle
&=& \left [ \frac{4}{3}\frac{(C_V^s+C_A^s)}{2}+\frac{1}{3}
\frac{(C_V^s-C_A^s)}{2} \right ] X
\eea
The values for $C_V^s$ and $C_A^s$ are taken as
\beq
C_V^s= -\frac{1}{2}+\frac{2}{3} \sin^2 \theta_W\;,
~~~~~~~~~C_A^s=-\frac{1}{2}\;.
\eeq
 Thus the amplitude for $B \to \phi K$ arising from the $Z$
mediated FCNC tree diagram is given as
\beq
A^{VLDQ}(\bar B^0 \to \phi \bar K_0)=\frac{G_F}{\sqrt 2}~
 \frac{4}{3} \left (-1 + \sin^2 \theta_W  
\right ) U_{sb} X\;.
\eeq

Using the experimental upper limit
$Br(B \to X_s l^+ l^-) < 4.2 \times 10^{-5}$ \cite{prl1}, in Ref.
\cite{ref12} the bound on
$|U_{bs}|$  is found to be $|U_{bs}| \leq 2
\times 10^{-3} $. Recently Belle
Collaboration \cite{prl2} has measured the branching
ratio for the process $B \to X_s l^+ l^-$  as
\beq
Br(B \to X_s l^+ l^-)=(6.1 \pm 1.4^{+1.4}_{-1.1}) \times 10^{-6}
\eeq
Using the above result one can obtain the value \cite{ref12,gb}
\beq
|Y_0(x_t)~\lambda_t^{bs}+C_{U2Z}~U_{bs}|=0.06\pm 0.03\;,\label{eq:lk1}
\eeq
where all the parameters in (\ref{eq:lk1}) are given in \cite{ref12}.
Thus one obtains the value of $U_{bs}$ as
\beq
|U_{bs}| \simeq 1 \times 10^{-3}
\eeq 
Now using  $\sin^2 \theta_W$=0.23, we find
\beq r_{NP}\simeq 0.58
\eeq
%%%%%%%%%%%%%%%%%%%%%%% FIGURE %%%%%%%%%%%%%%%%%%%%%%%%%%%%%%%

\begin{figure}[htb]
   \centerline{\epsfysize 4.0 truein \epsfbox{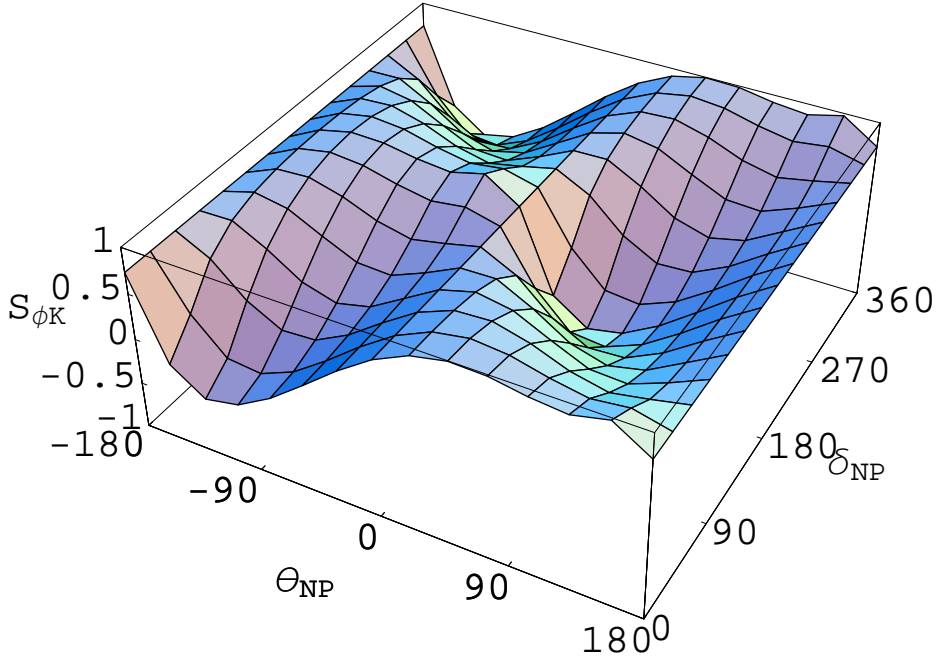}}
   \caption{
 3DPlot of $S_{\phi K}$ versus the weak phase
$\theta_{NP}$ and strong phase $\delta_{NP}$ (in degrees) for $r_{NP}=0.58$}
 \end{figure}

%%%%%%%%%%%%%%%%%%%%%%%%%%%%%%%%%%%%%%%%%%%%%%%%%%%%%%%%%%%%%%%%
%%%%%%%%%%%%%%%%%%%%%%% FIGURE %%%%%%%%%%%%%%%%%%%%%%%%%%%%%%%

\begin{figure}[htb]
   \centerline{\epsfysize 2.5 truein \epsfbox{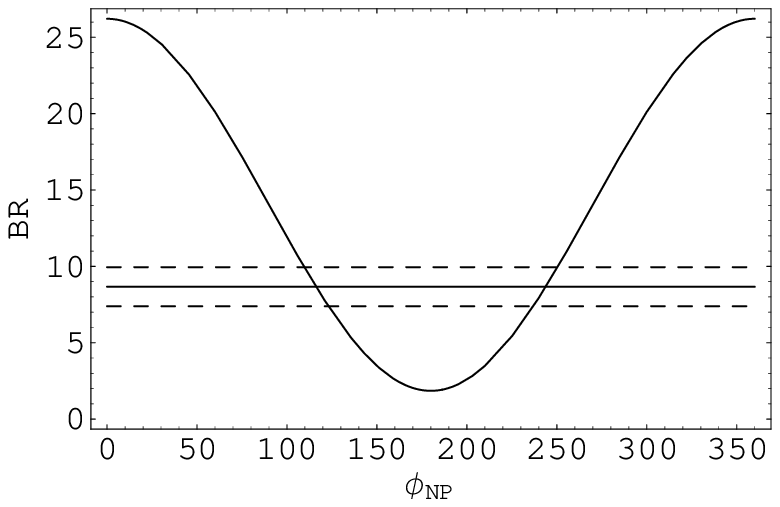}}
   \caption{
 Branching ratio of $B \to \phi K^0$ process 
(in units of $10^{-5}$) versus the phase
$\phi_{NP}$ (in degree).
The horizontal solid line is the central experimental
value whereas the dashed horizontal lines denote the error
limits.}
 \end{figure}

%%%%%%%%%%%%%%%%%%%%%%%%%%%%%%%%%%%%%%%%%%%%%%%%%%%%%%%%%%%%%%%%
The variation of $S_{\phi K}$ with respect to strong phase $\delta_{NP}$
and weak phase $\theta_{NP}$ according to Eq. (16)
in VLDQ model is shown in Figure-2 and the variation of branching ratio 
(15) is shown in Fig-3. It can
be seen from the figures that the observed asymmetry $S_{\phi K}$ and the
branching ratio
can be easily accommodated in this model.

\section{CP Violation in $ B \to \eta' K_S$ process}

At this stage we are in a position to test, as mentioned earlier, whether
the above two models (Model-III of THDM and VLDQ) can accommodate 
the result for
another similar mode, which seems to be in agreement with the SM.
In doing so, now we consider the $B^0 \to \eta' K^0 $ process.
\subsection{Contributions from SM and THDM}
In the SM, in addition to 
$b \to s \bar q q $ ( $q=(u,d,s)$) penguins, 
the $B^0 \to \eta' K_S $ process also receives a 
small contribution from 
color suppressed $ b \to u \bar u s $ tree diagram. We first find out the 
standard model contribution. The matrix element in the SM
is given as
\begin{eqnarray}
A(\bar B^0 \to \eta' \bar K^0 ) &=& \frac{G_F}{\sqrt 2}\biggr[V_{ub}V_{us}^* 
\sum_{i=1}^2
C_i^{eff}\langle \eta' \bar K^0 |O_i |\bar B^0\rangle \nonumber\\
&-&V_{tb}V_{ts}^* \sum_{i=3}^{10}
C_i^{eff}\langle \eta' \bar K^0 |O_i |\bar B^0\rangle
\biggr ]\;,
\end{eqnarray}
where $O_{1,2}$ are the tree and $O_{3-6(7-10)}$ are the QCD (electroweak)
penguin operators.
The matrix elements of these
operators are given  in the factorization approximation
 as \cite{khalil}
\bea
&& \langle \eta' \bar K^0|O_1 | \bar B^0 \rangle = \frac{1}{3} X_2~~~~~~
\langle \eta' \bar K^0|O_2 | \bar B^0 \rangle =  X_2 \nn\\
&& \langle \eta' \bar K^0|O_3 |\bar B^0 \rangle = \frac{1}{3} X_1
+2 X_2 + \frac{4}{3} X_3 \nn\\
&& \langle \eta' \bar K^0|O_4 |\bar B^0 \rangle =  X_1
+\frac{2}{3} X_2 + \frac{4}{3} X_3  \nn\\
&& \langle \eta' \bar K^0|O_5 | \bar B^0 \rangle = \frac{R_1}{3} X_1
-2 X_2 -\left (1- \frac{R_2}{3}\right )   X_3 \nn\\
&& \langle \eta' \bar K^0|O_6 | \bar B^0 \rangle = R_1 X_1
-\frac{2}{3} X_2 -\left ( \frac{1}{3}-R_2\right ) X_3
 \nn\\
&& \langle \eta' \bar K^0|O_7 | \bar B^0 \rangle = 
\frac{1}{2}\biggr[-\frac{R_1X_1}{3}- X_2
+\left (1-\frac{R_2}{3} \right )X_3 \biggr]\nn\\
&& \langle \eta' \bar K^0|O_8 |\bar B^0 \rangle =
\frac{1}{2}\biggr[-R_1X_1- \frac{X_2}{3}
+\left (\frac{1}{3} -R_2 \right )X_3 \biggr]\nn\\
&& \langle \eta' \bar K^0|O_9| \bar B^0 \rangle = 
\frac{1}{2}\biggr[-\frac{X_1}{3}+ X_2
-\frac{4}{3} X_3 \biggr]\nn\\
&& \langle \eta' \bar K^0|O_{10} |\bar B^0 \rangle = 
\frac{1}{2}\biggr[-X_1+ \frac{X_2}{3}
-\frac{4}{3}X_3 \biggr]
\eea
where
\bea
X_1 &=& i (m_B^2 - m_{\eta'}^2)F_0^{B \to \pi}(m_{K^0}^2 )\frac{X_{\eta'}}
{\sqrt 2} f_K \nn\\
X_2 &=&i (m_B^2 - m_{K^0}^2)F_0^{B \to K}(m_{\eta'}^2 )\frac{X_{\eta'}}
{\sqrt 2} f_{\pi} \nn\\
X_3 &=&i (m_B^2 - m_{K^0}^2)F_0^{B \to K}(m_{\eta'}^2 )Y_{\eta'}
\sqrt{ 2 f_{K}^2 -f_\pi^2} \nn\\
R_1 &=& \frac{2 m_{K^0}^2}{(m_b-m_d)(m_s+m_d)}\;,~~~~~~
R_2 = \frac{2(2m_{K^0}^2 -m_{\pi}^2}{(m_b-m_s)(m_s+m_s)}\;.
\eea
 $X_{\eta'}=0.57$ and $Y_{\eta'}=0.82$ are the mixing parameters of the
$u \bar u + d \bar d $ and $s \bar s$ components in the $\eta'$
meson \cite{khalil1}, which correspond to $\theta_P=-20^\circ$.
Thus the amplitude is given as
\bea
A(B^0 \to \eta'K^0) &=& i\frac{G_F}{2}\biggr[V_{ub}V_{us}^* a_2 X_2
-V_{tb}V_{ts}^*
\biggr\{
\left (a_4 -\frac{a_{10}}{2}+\left (a_6-\frac{a_8}{2} \right ) R_1 
\right )X_1 \nn\\
&+& \left (2(a_3-a_5) - \frac{1}{2}(a_7-a_9) \right ) X_2\nn\\
&+& \left (a_3 +a_4 -a_5 +\frac{1}{2}(a_7-a_9-a_{10})
+\left (a_6-\frac{a_8}{2} \right )R_2 \right )X_3 \biggr\}
\biggr]\label{eq:lm7}
\eea

The decay width can be given by
\beq
\Gamma ( B^0 \to \eta' K^0)= \frac{|\vec p|}{8 \pi m_B^2}|A(B^0
\to \eta' K^0)|^2
\eeq
Using $F_0^{(B \to \pi)}(m_{K^0}^2)=0.335$, $f_{K(\pi)}=0.16(0.13)$ GeV,
the quark masses as $(m_d,~ m_s, ~m_b)=
(0.0076,~ 0.122,~4.88)$ GeV and the values of the coefficients $a_i$'s
for $N_C=3$
from Ref. \cite{thdm2} we obtain the branching ratio in the 
standard model as
\beq
BR ( B^0 \to \eta'  K^0)|_{SM} =3.24\times 10^{-5}\;.
\eeq

which is slightly less than the current experimental
data \cite{pdg}
\beq
BR (  B^0 \to \eta'  K^0) =(5.8_{-1.3}^{+1.4})\times 10^{-5}\;.
\eeq

Now we consider the contributions arising from THDM. As discussed earlier
in this case due to the presence of new charged Higgs penguin diagrams,
the values of the effective Wilson coefficients $a_i$'s get modified.
Again substituting their values from \cite{thdm2} in Eq. (\ref{eq:lm7}), 
we obtain the transition amplitude as 
\beq
A^{THDM} (  B^0 \to \eta'  K^0)=A^{SM}(1+0.27~ e^{i(\theta_{NP}+
\delta_{NP})})\;.
\eeq
 Now taking
$r_{NP}=0.27$ and varying the weak phase $\theta_{NP}=\{-\pi,\pi\}
$ and strong phase  $\delta_{NP}=\{0,2\pi\} $ we can see from
Figure-4, that the observed
value of $S_{\eta' K}$ can be accommodated in the
THDM. Furthermore, the observed branching ratio can also
be explained in this model as seen from  from
Figure-5. If we take a crude assumption that the THDM and SM amplitudes 
interfere
constructively, the maximum value of branching ratio is found to be 
\beq
BR^{THDM} (  B^0 \to \eta'  K^0) =5.22\times 10^{-5}\;,
\eeq
which lies within the present experimental limits \cite{pdg}.

%%%%%%%%%%%%%%%%%%%%%%%%%%%%%%%%%%%%%5
\begin{figure}[htb]
   \centerline{\epsfysize 4.0 truein \epsfbox{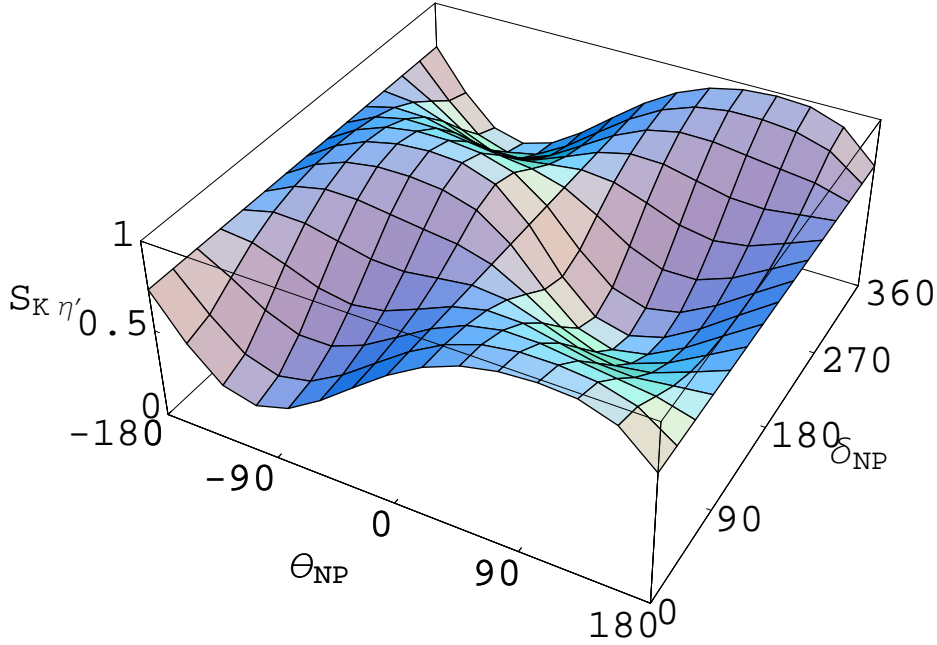}}
   \caption{
 3DPlot of $S_{\eta' K}$ versus the weak phase
$\theta_{NP}$ and strong phase $\delta_{NP}$ (in degrees) for $r_{NP}=0.27$}
 \end{figure}

%%%%%%%%%%%%%%%%%%%%%%%%%%%%%%%%%%%%%%%%%%%%%%%%%%%%%%%%%%%%%%%%
%%%%%%%%%%%%%%%%%%%%%%% FIGURE %%%%%%%%%%%%%%%%%%%%%%%%%%%%%%%

\begin{figure}[htb]
   \centerline{\epsfysize 2.5 truein \epsfbox{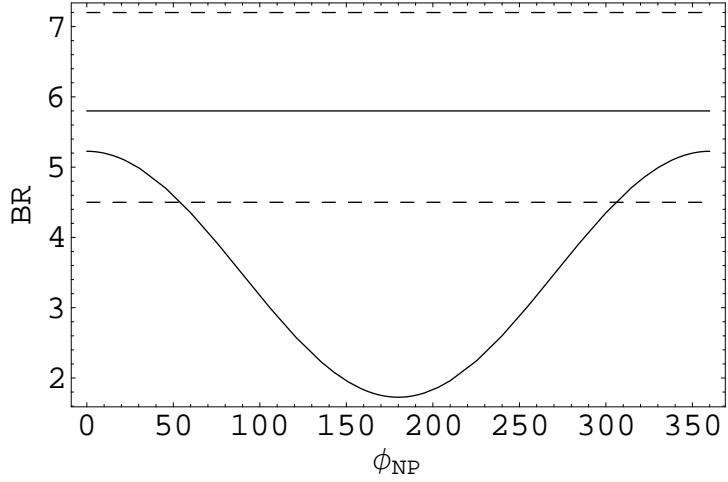}}
   \caption{
 Branching ratio of $B^0 \to \eta' K^0$ (in units of
$10^{-5}$) process in THDM versus the phase
$\phi_{NP}$ (in degree).
The horizontal solid line is the central experimental
value whereas the dashed horizontal lines denote the error
limits.}
 \end{figure}
 %%%%%%%%%%%%%%%%%%%%%%%%%%%%%%%%%%%%%%%%%%%%%

\subsection{ Contributions from VLDQ Model}

Now we consider the contributions arising from extra vector like 
down quark model. In this case the $B^0 \to \eta' K^0 $ process
proceeds through both color allowed and color suppressed tree level
$Z$-mediated FCNC diagrams. The corresponding operators are given as
\bea
&&O_1^{Z-FCNC}=[\bar s_\alpha \gamma^\mu(1-\gamma_5) b_\alpha]
[\bar q_\beta \gamma_\mu
(C_V^q-C_A^q \gamma_5)q_\beta] \nn\\
&&~~~~~~~~~\equiv
[\bar s_\alpha \gamma^\mu(1-\gamma_5) b_\alpha]\left [\bar q_\beta
 \gamma_\mu
\left \{ \frac{(C_V^q+C_A^q)}{2}(1-\gamma_5)
+ \frac{(C_V^q-C_A^q)}{2}(1+\gamma_5)\right \}q_\beta \right ]\nn\\
&&O_2^{Z-FCNC}=[\bar s_\beta \gamma^\mu(1-\gamma_5) b_\alpha]
[\bar q_\alpha \gamma_\mu
(C_V^q-C_A^q \gamma_5)q_\beta] \nn\\
&&~~~~~~~~\equiv
[\bar s_\beta \gamma^\mu(1-\gamma_5) b_\alpha][\bar q_\alpha
 \gamma_\mu
\left \{ \frac{(C_V^q+C_A^q)}{2}(1-\gamma_5)
+ \frac{(C_V^q-C_A^q)}{2}(1+\gamma_5) \right \}q_\beta ]
\eea

Using Fierz transformation and equation of motion the matrix 
elements of these operators are given as 
\bea
 \langle \eta' \bar K^0| O_1^{Z-FCNC}|\bar B^0 \rangle
 &=&(C_A^u  +C_A^d)X_2 +C_A^s X_3\nn\\
&+&\frac{1}{6}\biggr[\left (C_V^d(1+R_1)+C_A^d(1-R_1)\right )X_1
\nonumber\\
&+&\biggr ( C_V^s(1+R_2)+C_A^s(1-R_2) \biggr ) 
X_3 \biggr]
\eea
\bea 
 \langle \eta' \bar K^0| O_2^{Z-FCNC}| \bar B^0 \rangle & =&
\frac{1}{3}(C_A^u  +C_A^d)X_2 +
\frac{1}{3}C_A^s
 X_3\nn\\
&+&\frac{1}{2}\biggr[\left (C_V^d(1+R_1)+C_A^d(1-R_1)\right )X_1
\nn\\
&+& \biggr ( C_V^s(1+R_2)+C_A^s(1-R_2) 
\biggr ) X_3 \biggr]
\eea
So the amplitude for $B^0 \to  \eta' K^0  $ in VLDQ model is given as
\bea
A^{VLDQ}(\bar B^0  \to  \eta' \bar K^0) &=&  \frac{G_F}{\sqrt 2} U_{sb}
\biggr[ \frac{4}{3}\left\{(C_A^u+C_A^d) X_2 +C_A^s X_3\right \}\nn\\
&+&\frac{2}{3}\biggr\{\left (C_V^d(1+R_1) + C_A^d(1-R_1) \right )X_1\nn\\
&+& \biggr(C_V^s(1+R_2) + C_A^s(1-R_2) \biggr)X_3\biggr\}
\biggr]
\eea
Substituting the values of $C_{V(A)}^q $ as
\bea
&& C_V^u = \frac{1}{2}-\frac{4}{3}\sin^2 \theta_W\;,~~~~~~~~~
C_A^u=\frac{1}{2}\;,\nn\\
&& C_V^{(s,d)} = -\frac{1}{2}+\frac{2}{3}\sin^2 \theta_W\;,~~~~~
C_A^{(s,d)}=-\frac{1}{2}\;.
\eea
we find 
\beq
r_{NP} \simeq 0.72
\eeq
The variation of $S_{\eta' K^0}$ and the branching ratio according 
to Eqs. (16) and (15)in the
vector like down quark model are shown in Fig-6 and Fig-7. It can
be seen that the observed asymmetry and branching ratio for
 $B^0 \to \eta' K^0 $ mode
can be easily accommodated in this model.
\begin{figure}[htb]
   \centerline{\epsfysize 4.0 truein \epsfbox{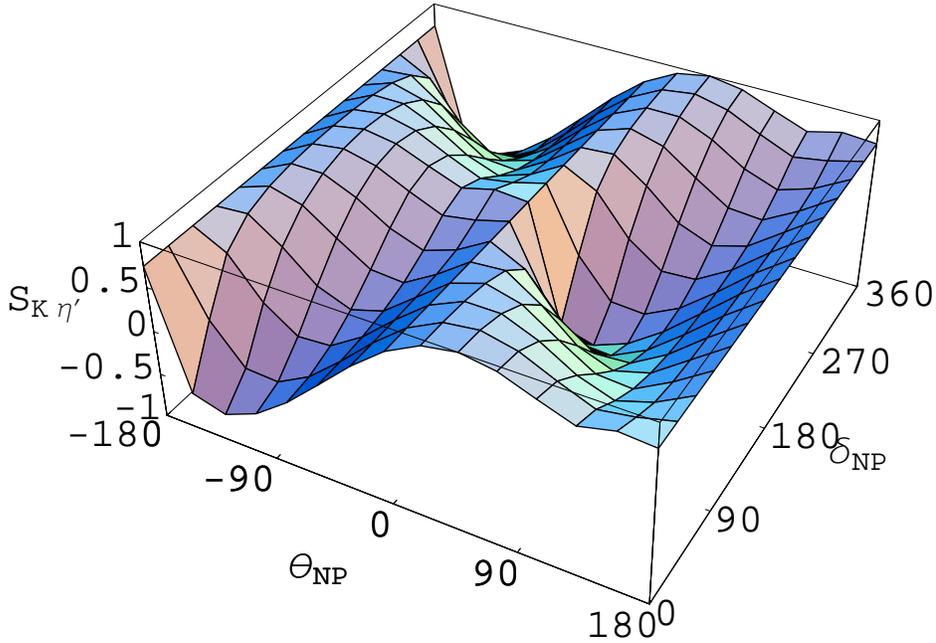}}
   \caption{
 3DPlot of $S_{\eta' K^0}$ versus the weak phase
$\theta_{NP}$ and strong phase $\delta_{NP}$ (in degrees) for $r_{NP}=0.72$}
 \end{figure}

%%%%%%%%%%%%%%%%%%%%%%%%%%%%%%%%%%%%%%%%%%%%%%%%%%%%%%%%%%%%%%%%
%%%%%%%%%%%%%%%%%%%%%%% FIGURE %%%%%%%%%%%%%%%%%%%%%%%%%%%%%%%

\begin{figure}[htb]
   \centerline{\epsfysize 2.5 truein \epsfbox{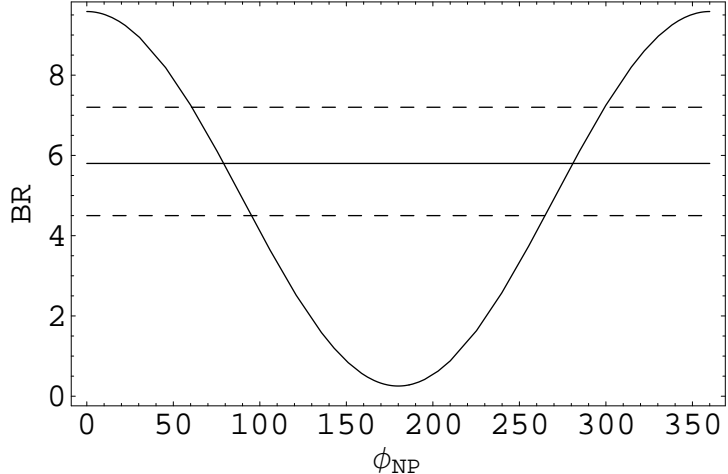}}
   \caption{
 Branching ratio of $B^0 \to \eta' K^0$ process 
(in units of $10^{-5}$) in VLDQ model versus the phase
$\phi_{NP}$ (in degree).
The horizontal solid line is the central experimental
value whereas the dashed horizontal lines denote the error
limits.}
 \end{figure}

\section{Conclusions}

To summarize, the time dependent CP asymmetry measurements in
$B\to \phi K_s$ gives $\sin 2\beta$, which is 2.7$\sigma$ deviation
from the corresponding value in $B \to \psi K_S$. According to
the SM expectation they should measure the same. Unlike the $B\to
\psi K_S$, which is a tree level process, $B \to \phi K_S$
occurs at one loop level, which allows room for new physics to play
an important role. In this paper, we have explored two simple beyond SM
scenarios, the two Higgs doublet model (model III) and
model with extra vector like down quark. We found that model III of THDM is
unable to explain, whereas vector like down quark model can
easily explain the result. 

It is important to note here that any new physics scenario that explains 
the $\phi K_s$ discrepancy must also explain another similar two-body decay 
$B \to \eta^\prime K_s$, which is also expected to give
the same value as of $ \psi K_s$ or $\phi K_s$, i.e., sin 2$\beta $.
In doing so it will be easy to rule out or narrow down the various NP
scenarios. We found that both the models (THDM-model-III and VLDQ) can explain
the $\eta^\prime K_s$ result. This in turn gives us the clue that the
VLDQ model may possibly be a strong contender for the
NP effects responsible in $B \to \phi K_s$. It is worthwhile
to emphasize that various supersymmetric models (as can be found in the 
literature) can explain the $\phi K_s$ discrepancy. But apart from
[13, 14] none of the scenarios so far explained the 
simultaneous explanation of
$\phi K_s$ and $\eta^\prime K_s$. On the other hand, our findings indicate 
that the simple non-supersymmetric extension of the 
SM in terms of the matter content should not be ignored for possible NP
candidature. Regardless of the sources of NP,
if in future the $\phi K_s$ result continues to be
different from the SM expectation, then it will certainly 
establish the presence of NP.

\section{Acknowledgements}
We thank Yuval Grossman for fruitful discussions. The work of RM
was supported in part by Department of Science and Technology,
Government of India through Grant No. SR/FTP/PS-50/2001.

%%%%%%%%%%%%%%%%%%%%% REFERENCES %%%%%%%%%%%%%%%%%%%%%%%%%%%%%%%%

\end{document}